\documentclass[5p,twocolumn]{elsarticle}
\usepackage{amssymb}
\usepackage{amsmath}
\usepackage{graphicx}
\usepackage{hyperref}
\usepackage{verbatim}

\journal{Computer Physics Communications}

\begin{document}

\title{QuTiP 2: A Python framework for the dynamics of open quantum systems}

\author[riken]{J. R. Johansson\fnref{fn1}\corref{cor1}}
\ead{robert@riken.jp}
\author[korea]{P. D. Nation\fnref{fn1}\corref{cor1}}
\ead{pnation@korea.ac.kr}
\author[riken,um]{Franco Nori}
\address[riken]{Advanced Science Institute, RIKEN, Wako-shi, Saitama 351-0198, Japan}
\address[korea]{Department of Physics, Korea University, Seoul 136-713, Korea}
\address[um]{Department of Physics, University of Michigan, Ann Arbor, Michigan 48109-1040, USA}
\fntext[fn1]{These authors contributed equally to this work.}
\cortext[cor1]{Corresponding authors}
\date{\today}

\begin{abstract}
We present version 2 of QuTiP, the Quantum Toolbox in Python. Compared to the preceding version [Comput. Phys. Comm. 183 (2012) 1760], we have introduced numerous new features, enhanced performance, made changes in the Application Programming Interface (API) for improved functionality and consistency within the package, as well as increased compatibility with existing conventions used in other scientific software packages for Python. The most significant new features include efficient solvers for arbitrary time-dependent Hamiltonians and collapse operators, support for the Floquet formalism, and new solvers for Bloch-Redfield and Floquet-Markov master equations.  Here we introduce these new features, demonstrate their use, and give a summary of the important backward-incompatible API changes introduced in this version. 
\end{abstract}

\maketitle
{\bf Program Summary}
\begin{small}
\noindent
\\
{\em Manuscript Title:} QuTiP 2: A Python framework for the dynamics of open quantum systems  \\
{\em Authors:} J. R. Johansson, P. D. Nation \\
{\em Program Title:} QuTiP: The Quantum Toolbox in Python   \\
{\em Journal Reference:}  \\
{\em Catalogue identifier:}   \\
{\em Licensing provisions:} GPLv3 \\
{\em Programming language:} Python   \\
{\em Computer:} i386, x86-64  \\
{\em Operating system:} Linux, Mac OSX \\
{\em RAM:} 2+~Gigabytes                                              \\
{\em Number of processors used:} 1+                              
{\em Classification:} 7 Condensed Matter and Surface Science   \\
{\em External routines/libraries:} NumPy, SciPy, Matplotlib, Cython \\
{\em Journal reference of previous version:} Comput. Phys. Comm. 183 (2012) 1760\\
{\em Does the new version supersede the previous version?:} Yes \\
{\em Nature of problem:} Dynamics of open quantum systems \\
{\em Solution method:} Numerical solutions to Lindblad, Floquet-Markov, and Bloch-Redfield master equations, as well as the Monte Carlo wave function method.\\
{\em Restrictions:} Problems must meet the criteria for using the master equation in Lindblad, Floquet-Markov, or Bloch-Redfield form.\\
{\em Running time:} A few seconds up to several tens of hours, depending on size of the underlying Hilbert space.\\
   \\
\end{small}

\section{Introduction}

The Quantum Toolbox in Python (QuTiP), is a generic framework for numerical simulation and computation of the dynamics of both open and closed quantum systems. This framework follows an object-oriented design that makes the programming of a quantum mechanical problem an intuitive process that closely follows the corresponding mathematical formulation. Although implemented in the interpreted programming language Python, the use of the NumPy and SciPy scientific libraries, and selective optimization using Cython, allows QuTiP to achieve performance that matches, or in many cases, exceeds that of natively compiled alternatives.  At the same time,  QuTiP provides a significantly more convenient and flexible programming environment that is easy to learn, and is well suited for use in the classroom.

Since the first major-version release of QuTiP \cite{johansson:2012} (the 1.x series), active development of the framework has resulted in a significant number of new features and performance enhancements, that have culminated in the second major-version release. Here we briefly describe the most significant changes and additional functionality introduced in this most recent release.

This paper is organized as follows.  In Sec.~\ref{sec:api-changes} we highlight the important API changes introduced in going from QuTiP 1.x to versions 2.x and higher. Section~\ref{sec:new} details the primary new features included in this latest version.  To illustrate the new functionality in QuTiP, Sec.~\ref{sec:examples} contains a selection of examples that highlight how these functions are used in numerical quantum simulations.  Finally, a list of all new user-accessible functions, including a brief description, is given in Appendix A.

\section{API changes}\label{sec:api-changes}

Here we list the backward-incompatible changes in the API of QuTiP 2 as compared to the previous version (1.1.4) described in Ref.~\cite{johansson:2012}. These changes are important when porting applications and simulations that are developed for QuTiP 1.1.4 to QuTiP 2.0 and higher. For newly developed simulations, we recommend following the documentation and examples for QuTiP 2.1 \cite{qutip}, in which case the following API changes are not relevant.  Note that this article covers version 2.1.0 of the QuTiP framework, and incorporates several features added since the initial 2.0 release.

\begin{enumerate}[1.]

\item
All quantum dynamics solvers (\texttt{mcsolve}, \texttt{mesolve}, \texttt{odesolve}, \texttt{essolve}, \texttt{brmesolve}, and \texttt{fmmesolve}) now return an \texttt{Odedata} instance, that contains all information about the solution (as opposed to data lists or Qobjs lists as in QuTiP version 1.1.4). A typical call to a time-evolution solver using the new API is:

\begin{footnotesize}
\begin{verbatim}
sol = solver(H, psi0, tlist, c_ops, e_ops)
\end{verbatim}
\end{footnotesize}

where the \texttt{sol.expect} or \texttt{sol.states} contain the lists of expectation values or \texttt{Qobj} instances, respectively, that would be returned by the same solver in QuTiP 1.1.4. With this new API, each solver can optionally store additional information in the return object such as, for example, the collapse times calculated in the Monte Carlo solver.

\item
The name of the function for Lindblad master equation solver has been changed from \texttt{odesolve} to \texttt{mesolve}.  The \texttt{odesolve} function can still be called, however it is officially deprecated, and will be removed in a future release.  Being a QuTiP version 1.x function, \texttt{odesolve} does not return an \texttt{Odedata} object.

\item
The order of the return values of the method \texttt{Qobj.eigenstates} have been swapped, so that the eigenenergies and eigenstates of a \texttt{Qobj} instance \texttt{op} are now returned in the following order:
    
\begin{footnotesize}
\begin{verbatim}
eigvals, eigkets = op.eigenstates()
\end{verbatim}
\end{footnotesize}

\item
Functions for calculating correlations using different solvers have now been consolidated under the functions
\texttt{correlation} and \texttt{correlation\_ss}, for transient and steady state correlations respectively.  Here, the selection of the underlying dynamics solver now is specified using the optional keyword argument \texttt{solve} that defaults to the Lindblad master equation (\texttt{mesolve}) if it is not explicitly specified.  For example:
\begin{footnotesize}
\begin{verbatim}
corr_mat = correlation(H, rho0, tlist, taulist, 
                       c_op, A, B, solver="me")
\end{verbatim}
\end{footnotesize}
where \texttt{solver} can be \texttt{"me"}, \texttt{"es"} or \texttt{"mc"}.

\end{enumerate}

\section{New features}\label{sec:new}

QuTiP 2 includes a wide variety of new computational functions, as well as utility functions for better handling of data.   Here we give a brief description the new major features in QuTiP 2.1. For full documentation of these new features, as well
as the rest of the QuTiP package, see the QuTiP 2.1 Documentation \cite{qutip}.  Examples illustrating the usage of these functions can be found in Sec.~\ref{sec:examples}.

\begin{itemize}
\item
  \textbf{Support for time-dependent collapse operators}: We have
  created a new system for representing time-dependent quantum
  operators used in defining system Hamiltonians and collapse operators for
  the Lindblad master equation and the Monte Carlo solvers. This
  allows support for arbitrary time-dependent collapse operators (the new
  method is still backwards compatible with the python function callback
  method used for time-dependent Hamiltonians in QuTiP version 1.1.4).
  This new method of defining arbitrary time-dependencies is both more
  efficient and more flexible, allowing for high performance simulations
  of arbitrary time-dependent quantum systems.  In particular, many problems of interest
  may be compiled at runtime into C code via Cython \cite{behnel:2011}.  This particular feature,
  and its implementation, will be discussed elsewhere \cite{johansson:2013}.

\item
  \textbf{Floquet formalism, Floquet-Markov master equation}: For periodic
  time-dependent systems, the Floquet formalism can be a useful
  technique where the original time-dependent problem is transformed into a
  time-independent problem using the time-dependent Floquet-modes as
  the basis set. In QuTiP 2.0 we added a new module for the Floquet-related
  decomposition of time-dependent problems, and the evolution of unitary
  and dissipative dynamics using equations of motion and master
  equations in the Floquet formalism.

\item
  \textbf{Bloch-Redfield master equation solver:} A new quantum dynamics
  solver for the time evolution according to the Bloch-Redfield master
  equation is now included in QuTiP. While not as efficient as the Lindblad master
  equation solver, in situations where the environment is expressed in terms of its noise power
  spectrum, rather than phenomenological decay and dephasing rates used in the Lindblad formalism, the Bloch-Redfield master equation has significant advantages.

\item
  \textbf{Quantum Process Tomography:} Quantum process tomography (QPT) \cite{mohseni:2008} is a useful technique for characterizing experimental implementations of quantum gates involving a small number of qubits. It can also be a useful theoretical tool that gives insight into how a given process transforms density matrices, and can be used, for example, to study how noise or other imperfections deteriorate quantum gate operations. Unlike the fidelity or trace distance, that give a single number indicating how far from ideal a gate is, quantum process tomography gives detailed information as to exactly what kind of errors various imperfections and losses introduce.

\item
  \textbf{Functions for generating random states and matrices}: It is now possible to generate
random kets, density matrices, Hamiltonians, and Unitary operators.  This includes the ability to set
the sparsity (density) of the resultant quantum object.

\item
  \textbf{Support for sparse eigensolvers}: The quantum object (Qobj) methods
\texttt{eigenstates}, \texttt{eigenenergies}, and \texttt{groundstate} can now use sparse eigensolvers for systems
with large Hilbert spaces.  This option is not enabled by default, and must be set with the keyword argument \texttt{sparse}.  
  
\item
  \textbf{New entropy and entanglement functions}: Functions for calculating the concurrence, mutual information,
  and conditional entropy have been added.

\item
  \textbf{New operator norms}: When calculating operator norms, one can now select between the following norms: trace, Frobius, one, and max.  The trace norm is chosen by default.  For ket and bra vectors, only the L2-norm is available.

\item
  \textbf{Saving and loading data}: Saving and loading of quantum objects and array data is now internally supported by QuTiP.  The storage and retrieval of all quantum objects and Odedata objects can be accomplished via the \texttt{qsave} and \texttt{qload} functions, respectively.  In order to facilitate the export of QuTiP data to other programs, the \texttt{file\_data\_store} and \texttt{file\_data\_read} allow the user to read and write array data, both real and complex, into text files with a wide variety of formatting options.

\item
  \textbf{Performance improvements}: In QuTiP 2.1, numerous performance
  optimizations have been implemented, including more efficient quantum
  object creation, significantly faster \texttt{ptrace} implementation, and an improved
  \texttt{steadystate} solver.

\item
  \textbf{Unit tests for verification of installation}: The installation
  of QuTiP 2.1 comes with a set of unit tests that can be used to verify that the installation was successful, 
and that the underlying routines are functioning as expected.

\end{itemize}

\section{Example scripts featuring new functionality}\label{sec:examples}
In this section we highlight, via examples, several of the main features added in QuTiP 2.1 and listed in Sec.~\ref{sec:new}.  Although we will demonstrate the use of the new time-dependent evolution framework, a full discussion of this feature is presented elsewhere \cite{johansson:2013}.  The examples listed below, as well as a growing collection of additional demonstrations, can be found on the QuTiP website \cite{qutip}, or run after installing QuTiP using the \texttt{demos} function.  For brevity, we do not include the portions of code related to figure generation using the matplotlib framework \cite{matplotlib}.

\subsection{API changes to dynamics solvers}\label{sec:ex-api}
Here we demonstrate using the new \texttt{Odedata} class that is returned by the \texttt{mcsolve}, \texttt{mesolve}, \texttt{brmesolve}, and \texttt{fmmesolve} evolution solvers in QuTiP version 2.1.  To better illustrate the API changes, we have recoded the two-qubit gate example from Ref.~\cite{johansson:2012} Sec.~(4.1) that is written using the older QuTiP 1.x API.  The sections of the script featuring the new API are indicated below:

\begin{footnotesize}
\begin{verbatim}
from qutip import *

g  = 1.0 * 2 * pi   # coupling strength
g1 = 0.75           # relaxation rate
g2 = 0.05           # dephasing rate
n_th = 0.75         # bath avg. thermal excitations
T = pi/(4*g)        # gate period

# construct Hamiltonian
H = g * (tensor(sigmax(), sigmax()) +
         tensor(sigmay(), sigmay()))
# construct inital state
psi0 = tensor(basis(2,1), basis(2,0))     

# construct collapse operators
c_ops = []
## qubit 1 collapse operators
sm1 = tensor(sigmam(), qeye(2))
sz1 = tensor(sigmaz(), qeye(2))
c_ops.append(sqrt(g1 * (1+n_th)) * sm1)
c_ops.append(sqrt(g1 * n_th) * sm1.dag())
c_ops.append(sqrt(g2) * sz1)
## qubit 2 collapse operators
sm2 = tensor(qeye(2), sigmam())
sz2 = tensor(qeye(2), sigmaz())
c_ops.append(sqrt(g1 * (1+n_th)) * sm2)
c_ops.append(sqrt(g1 * n_th) * sm2.dag())
c_ops.append(sqrt(g2) * sz2)

# evolve the dissipative system
tlist = linspace(0, T, 100)
medata  = mesolve(H, psi0, tlist, c_ops, [])

## NEW API CALL ##
# extract density matrices from Odedata object
rho_list = medata.states

# get final density matrix for fidelity comparison
rho_final = rho_list[-1]
# calculate expectation values 
n1 = expect(sm1.dag() * sm1, rho_list)
n2 = expect(sm2.dag() * sm2, rho_list)     
# calculate the ideal evolution 
medata_ideal = mesolve(H, psi0, tlist, [], [])

## NEW API CALL ##
# extract states from Odedata object
psi_list = medata_ideal.states

# calculate expectation values 
n1_ideal = expect(sm1.dag() * sm1, psi_list)
n2_ideal = expect(sm2.dag() * sm2, psi_list)
# get last ket vector for comparison
psi_ideal = psi_list[-1]
# output is ket since no collapse operators.
rho_ideal = ket2dm(psi_ideal)

# calculate the fidelity of final states
F = fidelity(rho_ideal, rho_final)

\end{verbatim}
\end{footnotesize}

\subsection{Floquet modes of a driven two-level system}

\begin{figure}[htb]
\begin{center}
\includegraphics[width=9cm]{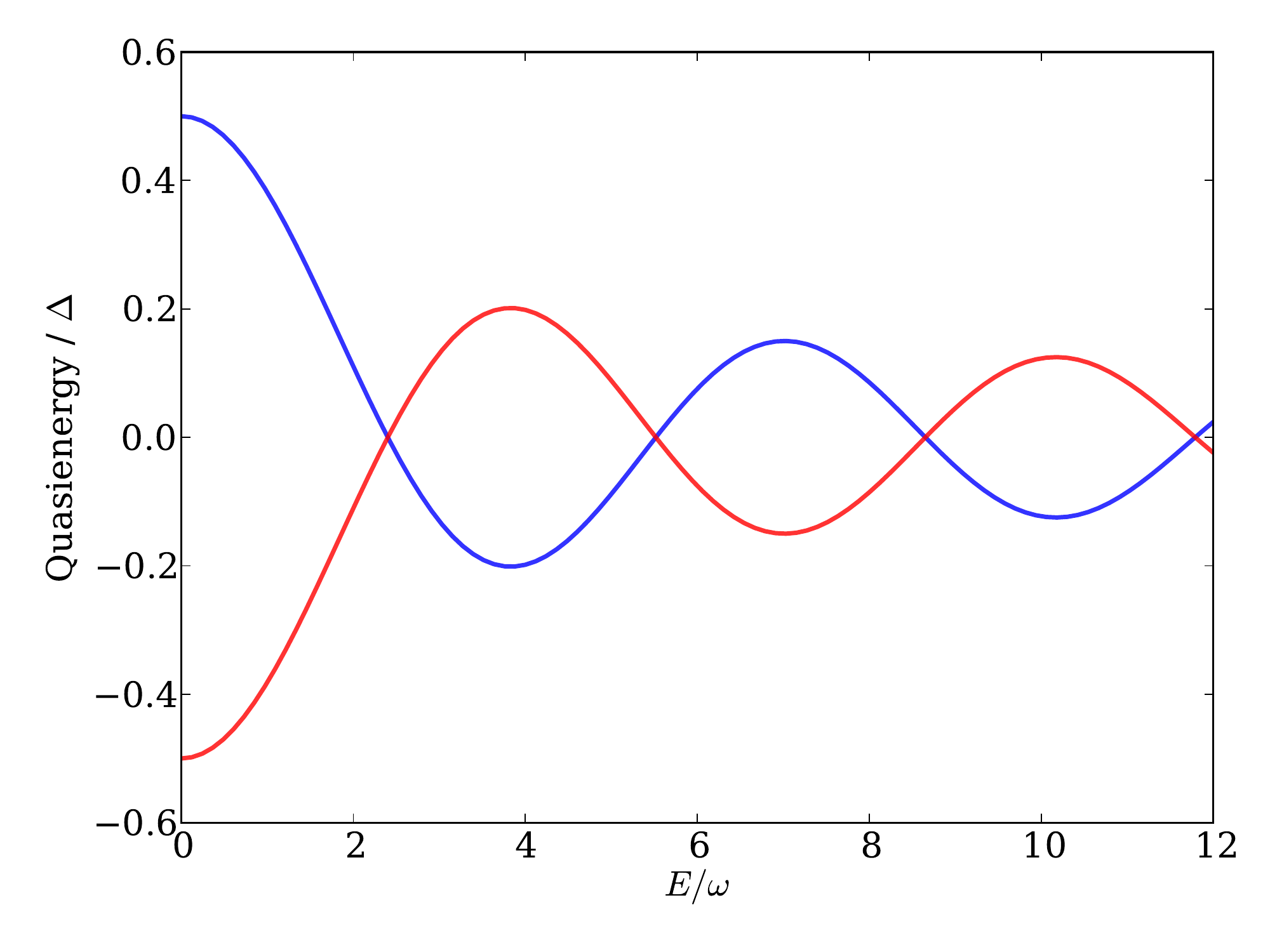}
\caption{(Color) Quasienergies corresponding to the two Floquet basis states of a driven two-level system as the driving strength is increased.  Here the quasienergy and driving amplitude are expressed in units of the qubit energy splitting and driving frequency ($\hbar=1$),  respectively.  In this simulation, $\Delta=1\times 2\pi$ and $\omega=8\times 2\pi$.}
\label{fig:quasi}
\end{center}
\end{figure}

Following the example in Ref.~ \cite{creffield:2003}, here we calculate the quasienergies for the time-dependent Floquet basis vectors of a sinusoidally driven two-level system \cite{shevchenko:2010} with Hamiltonian
\begin{equation}
H=\frac{\Delta}{2}\sigma_{z}+\frac{E}{2}\cos\left(\omega t\right)\sigma_{x},
\end{equation}
where $\Delta$ is the qubit energy splitting and $\omega$ is the driving frequency, for different values of the driving amplitude $E$.  The results of the simulation are presented in Fig.~\ref{fig:quasi} .
\begin{footnotesize}
\begin{verbatim}
from qutip import *

delta = 1.0 * 2 * pi  # bare qubit sigma_z coefficient
omega = 8.0 * 2 * pi  # driving frequency
T     = (2*pi)/omega  # driving period

# vector of driving amplitudes
E_vec = linspace(0.0, 12.0, 100) * omega

# generate spin operators
sx = sigmax()
sz = sigmaz()

# create array for storing energy values
q_energies = zeros((len(E_vec), 2))

# define time-independent Hamiltonian term
H0 = delta/2.0 * sz 
args = {'w': omega}

# loop over driving amplitudes
for idx, E in enumerate(E_vec):
    # amplitude-dependent Hamiltonian term
    H1 = E/2.0 * sx
    
    # H = H0 + H1 * cos(w * t) in 'list-string' format
    H = [H0, [H1, 'cos(w * t)']]    
        
    # find the Floquet modes
    f_modes, f_energies = floquet_modes(H, T, args)
    # record quasi-energies
    q_energies[idx,:] = f_energies

\end{verbatim}
\end{footnotesize}

\subsection{Floquet Evolution}\label{sec:ex-floquet}
\begin{figure}[htb]
\begin{center}
\includegraphics[width=9cm]{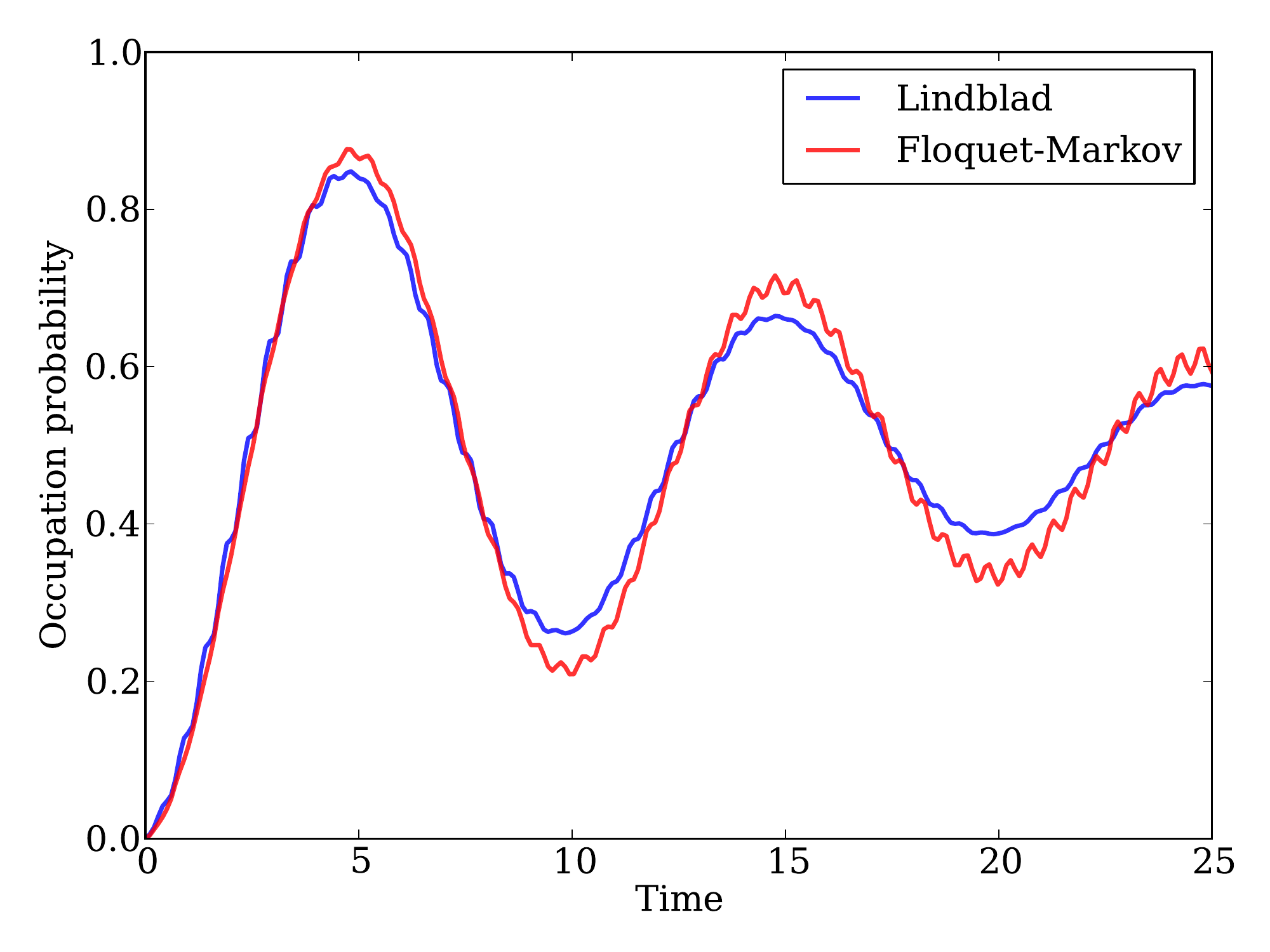}
\caption{(Color) Occupation probability of a sinusoidally driven qubit, initially in its ground state, under both Lindblad and Floquet-Markov master equation evolution, where the qubit parameters are $\Delta=0$, $\epsilon=1.0\times 2\pi$, and the relaxation and dephasing rates are given by $\gamma_{1}=0.05$, $\gamma_{2}=0$.  The driving term has amplitude $A=0.1\times 2\pi$ and frequency $\omega=1.0\times 2\pi$.  Here, the spectral noise density of the environment is assumed to be Ohmic. }
\label{fig:floquet}
\end{center}
\end{figure}

A driven system that is interacting with its environment is not necessarily well described by the standard Lindblad master equation as its dissipation process could be time-dependent due to the driving. In such cases a rigorous approach would be to take the driving into account when deriving the master equation. This can be done in many different ways, but one common approach is to derive the master equation in the Floquet basis, the Floquet-Markov master equation \cite{grifoni:1998}.  In QuTiP, this Floquet-Markov master equation is implemented in the \texttt{fmmesolve} function.  As this approach is for time-dependent systems, here we model a sinusoidally driven qubit with Hamiltonian
\begin{equation}
H=-\frac{\Delta}{2}\sigma_{x}-\frac{\epsilon}{2}\sigma_{z}-A\sigma_{x}\sin\omega t
\end{equation}
where $\Delta$ and $\epsilon$ are the coupling and energy splitting constants, while $A$ and $\omega$ are the driving strength and frequency, respectively.  In addition, we define the spectral density of the environmental noise to be Ohmic.  In Fig.~\ref{fig:floquet} we plot the occupation probability of the qubit for both the Lindblad and Floquet-Markov master equations as a function of time.

\begin{footnotesize}
\begin{verbatim}
from qutip import *

gamma1 = 0.05         # relaxation rate
gamma2 = 0.0          # dephasing  rate
delta = 0.0 * 2 * pi  # qubit sigma_x coefficient
eps0  = 1.0 * 2 * pi  # qubit sigma_z coefficient
A     = 0.1 * 2 * pi  # driving amplitude
w     = 1.0 * 2 * pi  # driving frequency
T     = 2*pi / w      # driving period
psi0  = basis(2,0)    # initial state
tlist = linspace(0, 25.0, 250)

def J_cb(omega):
    """ Noise spectral density """
    return 0.5 * gamma1 * omega/(2*pi)

# Hamiltonian in list-string format
args = {'w': w}
H0 = - delta/2.0 * sigmax() - eps0/2.0 * sigmaz()
H1 = - A * sigmax()
H = [H0, [H1, 'sin(w * t)']]

# -----------------------------------------------------
# Lindblad equation with time-dependent Hamiltonian
# 
c_ops = [sqrt(gamma1) * sigmax(), 
         sqrt(gamma2) * sigmaz()]
p_ex_me = mesolve(H, psi0, tlist, c_ops, 
                  [num(2)], args=args).expect[0]
 
# -----------------------------------------------------
# Floquet-Markov master equation dynamics
#       
rhs_clear() # clears previous time-dependent Hamiltonian

# find initial Floquet modes and quasienergies
f_modes_0, f_energies = floquet_modes(H, T, args, False)    
       
# precalculate Floquet modes for the first driving period
f_modes_table = floquet_modes_table(f_modes_0, f_energies,
                    linspace(0, T, 500+1), H, T, args) 

# solve the Floquet-Markov master equation
rho_list = fmmesolve(H, psi0, tlist, [sigmax()]
                     [], [J_cb], T, args).states

# calculate expectation values in the computational basis
p_ex_fmme = zeros(shape(p_ex_me))
for idx, t in enumerate(tlist):
    f_modes_t = floquet_modes_t_lookup(f_modes_table, t, T) 
    p_ex_fmme[idx] = expect(num(2), 
                rho_list[idx].transform(f_modes_t, False))
\end{verbatim}
\end{footnotesize}

\subsection{Bloch-Redfield master equation}\label{sec:ex-brme}
\begin{figure}[htb]
\begin{center}
\includegraphics[width=8cm]{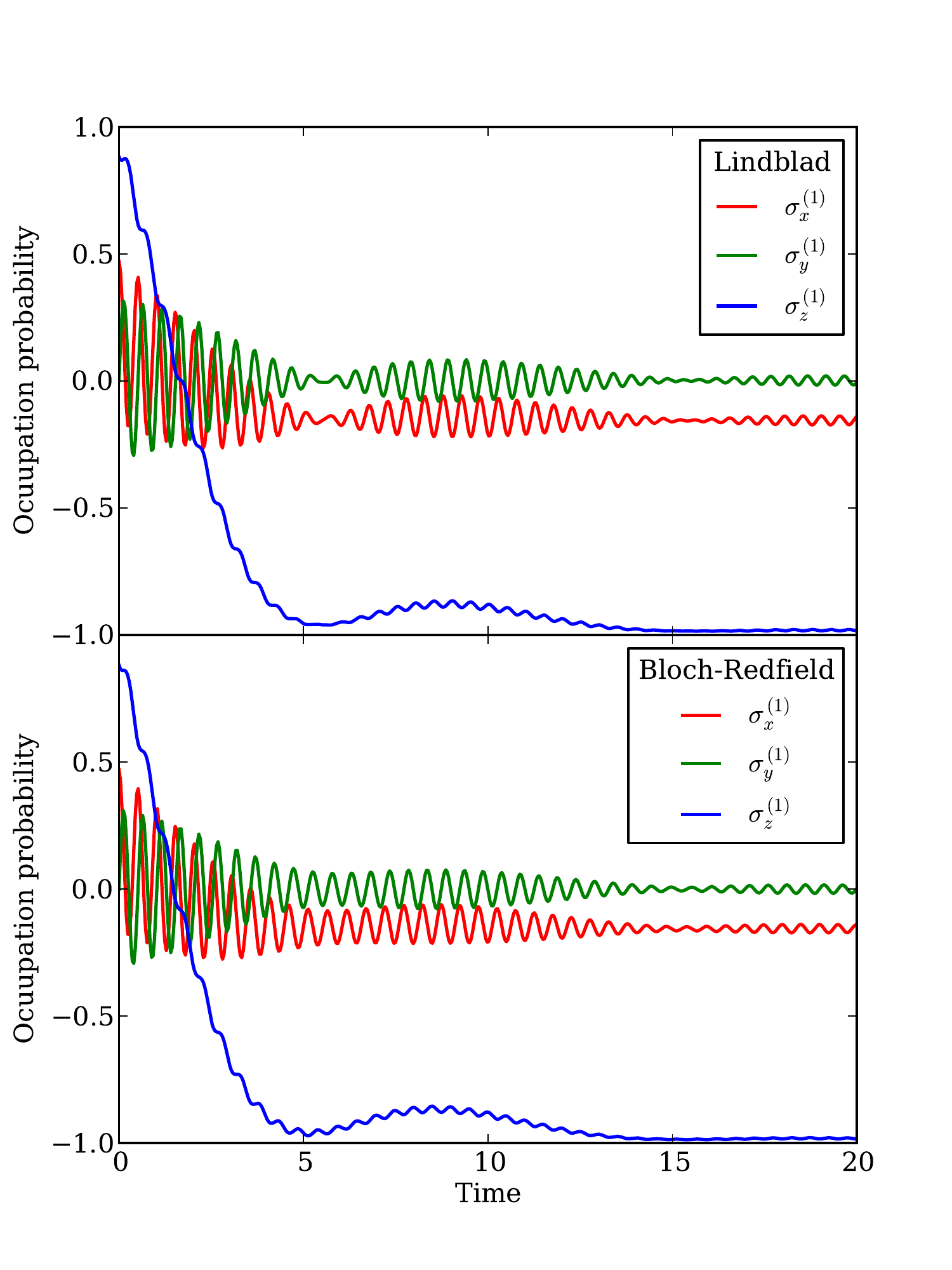}
\caption{(Color) Expectation values for the spin operators of qubit \#1, for both the Lindblad master equation (top) and the Bloch-Redfield master equation (bottom), where the qubit environment in the latter case is assumed to have an Ohmic spectrum.  Here, $\omega_{1}=\omega_{2}=1.0\times 2\pi$, $\theta_{1}=0.025\times 2\pi$, $\theta_{2}=0$, and the qubit relaxation terms for qubit \#1 and qubit \#2 are give by $\gamma_{1}^{(1)}=0.25$ and $\gamma_{1}^{(2)}=0.35$, respectively.  The qubit coupling is $g=0.05\times 2\pi$.  In this simulation, the dephasing terms are assumed to be zero.}
\label{fig:brmesolve}
\end{center}
\end{figure}

The Lindblad master equation is constructed so that it describes a physical evolution of the density matrix (i.e., trace and positivity preserving), but it does not provide a connection to any underlying microscopic physical model. However, a microscopic model can in some cases be advantagous, as for example in systems with varying energy biases and eigenstates that couple to an environment in some well-defined manner, through a physically motivated system-environment interaction operator that can be related to a noise-power spectrum.  The Bloch-Redfield formalism is one such approach to derive a master equation from the underlying microscopic physics of the system-bath coupling.  To highlight the differences inherent in these two approaches, in Fig.~\ref{fig:brmesolve} we plot the expectation values for the spin-operators of the first qubit in a coupled qubit system given by the Hamiltonian
\begin{align}\label{eq:blochH}
H&=\omega_{1}\left[\cos\theta_{1}\sigma_{z}^{(1)}+\sin\theta_{1}\sigma_{x}^{(1)}\right]\\
&+\omega_{2}\left[\cos\theta_{2}\sigma_{z}^{(2)}+\sin\theta_{2}\sigma_{x}^{(2)}\right]+g\sigma_{x}^{(1)}\sigma_{x}^{(2)} \nonumber
\end{align}
with the initial state $\left|\phi\rangle\right.=\left|\psi_{1}\rangle\right.\left|\psi_{2}\rangle\right.$ with
\begin{align*}
\left|\psi_{1}\rangle\right.&=\frac{5}{\sqrt{17}}\left[0.8\left|0\rangle_{(1)}\right.+(1-0.8)\left|1\rangle_{(1)}\right.\right] \\
\left|\psi_{2}\rangle\right.&=\frac{5}{\sqrt{17}}\left[(1-0.8)\left|0\rangle_{(2)}\right.+0.8\left|1\rangle_{(2)}\right.\right]. 
\end{align*}
where the subscripts indicate which qubit the state belongs to.  In Eq.~(\ref{eq:blochH}),  $g$ is the qubit coupling, $\omega_{1}$ and $\omega_{2}$ are the qubit frequencies, and finally $\theta_{1}$ and $\theta_{2}$ represent the angles of each qubit with respect to the $\sigma_{z}$ direction.  In this example, the qubit environments in the Bloch-Redfield simulation are assumed to have Ohmic spectrum.  The code for the corresponding simulation is given below:

\begin{footnotesize}
\begin{verbatim}
from qutip import *

w     = array([1.0,1.0])*2*pi    # qubit angular frequency
theta = array([0.025,0.0])*2*pi  # angle from sigma_z axis
gamma1 = [0.25, 0.35]            # qubit relaxation rate
gamma2 = [0.0, 0.0]              # qubit dephasing rate
g      = 0.1 * 2 * pi            # coupling strength
# initial state
a = 0.8
psi1 = (a*basis(2,0)+(1-a)*basis(2,1)).unit()
psi2 = ((1-a)*basis(2,0)+a*basis(2,1)).unit()
psi0 = tensor(psi1, psi2)
# times at which to evaluate expectation values
tlist = linspace(0, 15, 500)

# operators for qubit 1
sx1 = tensor(sigmax(), qeye(2))
sy1 = tensor(sigmay(), qeye(2))
sz1 = tensor(sigmaz(), qeye(2))
sm1 = tensor(sigmam(), qeye(2))
# operators for qubit 2
sx2 = tensor(qeye(2), sigmax())
sy2 = tensor(qeye(2), sigmay())
sz2 = tensor(qeye(2), sigmaz())
sm2 = tensor(qeye(2), sigmam())
# Hamiltonian
# qubit 1
H  = w[0] * (cos(theta[0]) * sz1 + sin(theta[0]) * sx1)
# qubit 2
H += w[1] * (cos(theta[1]) * sz2 + sin(theta[1]) * sx2)
# interaction term
H += g * sx1 * sx2

# -----------------------------------------------------
# Lindblad master equation
#
c_ops = []
c_ops.append(sqrt(gamma1[0]) * sm1)
c_ops.append(sqrt(gamma1[1]) * sm2)
    
lme_results = mesolve(H, psi0, tlist, c_ops,
                      [sx1, sy1, sz1])

# -----------------------------------------------------
# Bloch-Redfield master equation
#  
def ohmic_spectrum1(w):
    if w == 0.0:
        # dephasing inducing noise
        return 0.5 * gamma2[0]
    else:
        # relaxation inducing noise
        return 0.5 * gamma1[0]*w/(2*pi)*(w > 0.0)

def ohmic_spectrum2(w):
    if w == 0.0:
        # dephasing inducing noise
        return 0.5 * gamma2[1]
    else:
        # relaxation inducing noise
        return 0.5 * gamma1[1]*w/(2*pi)*(w > 0.0)

brme_results = brmesolve(H, psi0, tlist, [sx1, sx2],
                 [sx1, sy1, sz1], [ohmic_spectrum1,
                 ohmic_spectrum2])
\end{verbatim}
\end{footnotesize}

\subsection{Quantum process tomography}\label{sec:ex-qpt}
\begin{figure}[htb]
\begin{center}
\includegraphics[width=8cm]{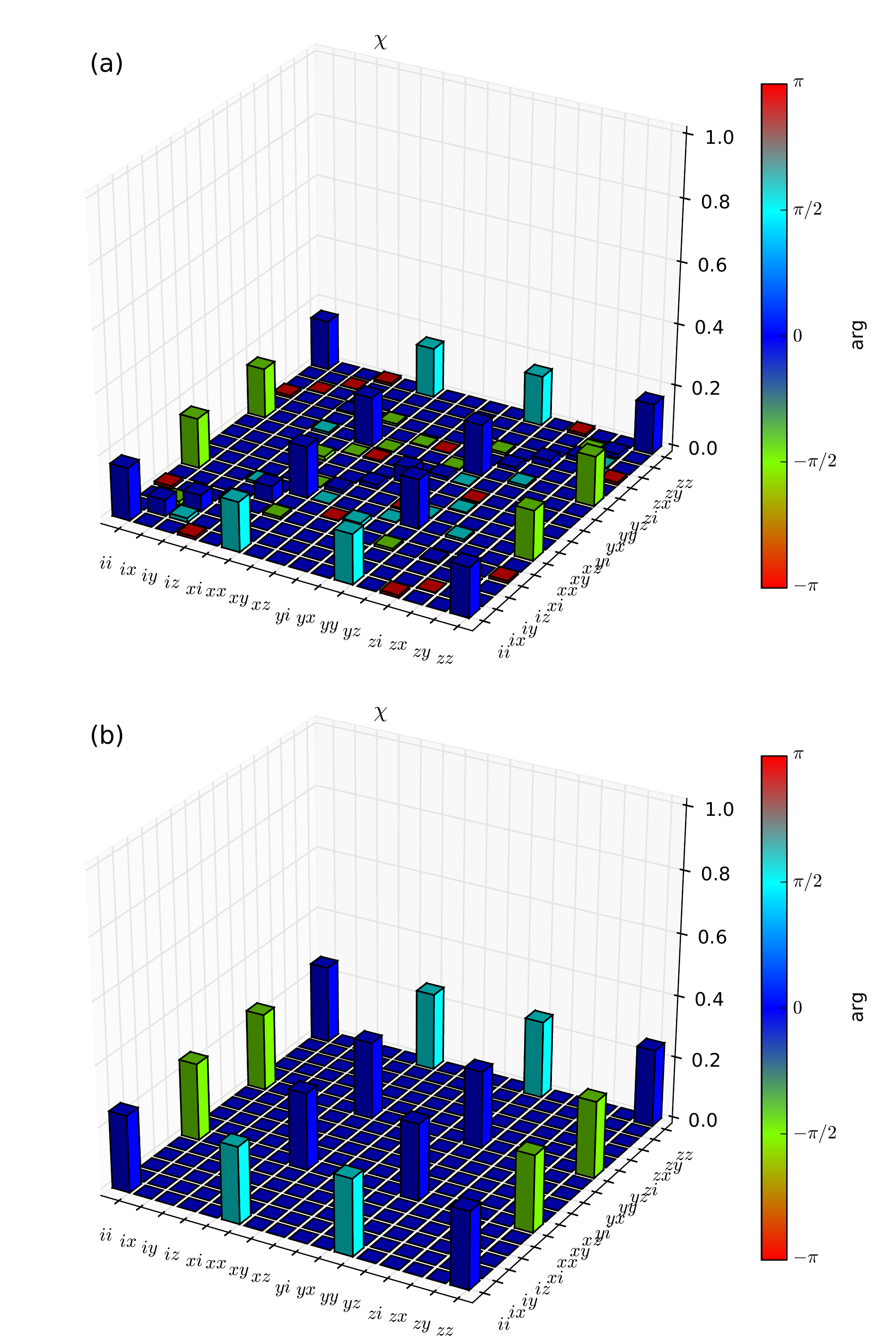}
\caption{(Color) (a) The QPT process $\chi$-matrix for the dissipative iSWAP gate between two-qubits. Here the color indicates the phase of each matrix element. The qubit-qubit coupling strength is $g=1.0\times 2\pi$, whereas the relaxation and dephasing rates are $g1=0.75$ and $g2=0.25$, respectively.  (b) The ideal iSWAP gate $\chi$-matrix when the qubit dissipation and dephasing are not present.}
\label{fig:qpt}
\end{center}
\end{figure}

To demonstrate quantum process tomography, here we simulate the effects of relaxation and dephasing on the two-qubit iSWAP gate \cite{schuch:2003} when the qubits are coupled to a thermal environment with on average $\langle n\rangle=1.5$ excitations.  The $\chi$-matrix obtained from QPT contains all the information about the dynamics of this open quantum system.  In Fig. \ref{fig:qpt} we plot the $\chi$-matrix for both the dissipative and corresponding ideal (unitary) iSWAP gate dynamics. 

\begin{footnotesize}
\begin{verbatim}
from qutip import *

g = 1.0 * 2 * pi  # coupling strength
g1 = 0.75         # relaxation rate
g2 = 0.25         # dephasing rate
n_th = 1.5        # bath avg. thermal excitations
T = pi/(4*g)      # gate period

H = g*(tensor(sigmax(),sigmax())+tensor(sigmay(),sigmay()))
psi0 = tensor(basis(2,1), basis(2,0))

c_ops = []
# qubit 1 collapse operators
sm1 = tensor(sigmam(), qeye(2))
sz1 = tensor(sigmaz(), qeye(2))
c_ops.append(sqrt(g1 * (1+n_th)) * sm1)
c_ops.append(sqrt(g1 * n_th) * sm1.dag())
c_ops.append(sqrt(g2) * sz1)
# qubit 2 collapse operators
sm2 = tensor(qeye(2), sigmam())
sz2 = tensor(qeye(2), sigmaz())
c_ops.append(sqrt(g1 * (1+n_th)) * sm2)
c_ops.append(sqrt(g1 * n_th) * sm2.dag())
c_ops.append(sqrt(g2) * sz2)

# process tomography basis
op_basis = [[qeye(2), sigmax(), sigmay(), sigmaz()]] * 2
op_label = [["i", "x", "y", "z"]] * 2

# dissipative gate
U_diss = propagator(H, T, c_ops)
chi = qpt(U_diss, op_basis)
qpt_plot_combined(chi, op_label)

# ideal gate 
U_psi = (-1j * H * T).expm()
U_ideal = spre(U_psi) * spost(U_psi.dag())
chi = qpt(U_ideal, op_basis)
qpt_plot_combined(chi, op_label)
\end{verbatim}
\end{footnotesize}

\subsection{Exporting QuTiP data}\label{sec:ex-export}
\begin{figure}[htb]
\begin{center}
\includegraphics[width=8cm]{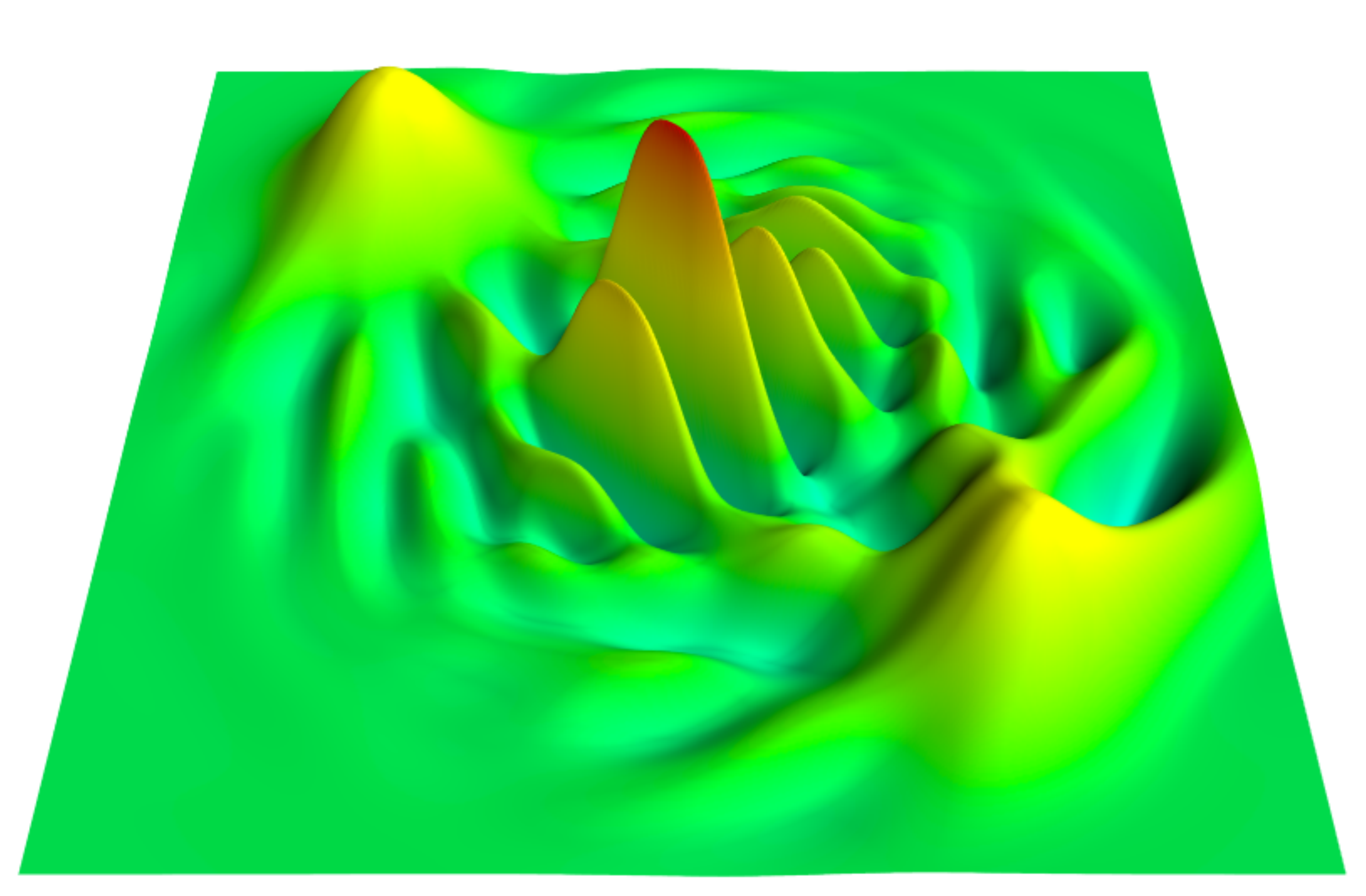}
\caption{(Color) Wigner function for the state $\left|\Psi\rangle\right.=\left|\alpha\rangle\right.+\left|-\alpha\rangle\right.+|\tilde{\phi}\rangle$ where $\alpha=2+2j$ is the coherent state amplitude, and $|\tilde{\phi}\rangle$ is a randomly generated state vector.  This plot is generated in Mayavi using data from the QuTiP \texttt{wigner} function.}
\label{fig:wigner}
\end{center}
\end{figure}

Finally, we demonstrate the exporting of data generated in QuTiP to an external plotting program using the \texttt{file\_data\_store} and \texttt{file\_data\_read} to save and load the data, respectively.  To keep the example completely in Python, we have chosen to use Mayavi \cite{mayavi} to plot a Wigner function in Fig.~\ref{fig:wigner} generated in QuTiP corresponding to the state $\left|\Psi\rangle\right.=\left|\alpha\rangle\right.+\left|-\alpha\rangle\right.+|\tilde{\phi}\rangle$ that is a Schr\"{o}dinger cat state  consisting of two coherent states with complex amplitude $\alpha$, with an additional random ket vector $|\tilde{\phi}\rangle$ created using QuTiP's random state generator:

\begin{footnotesize}
\begin{verbatim}
from qutip import *
# Number of basis states
N = 20 
# amplitude of coherent states
alpha = 2.0 + 2.0j
# define ladder operators
a = destroy(N)
# define displacement operators
D1 = displace(N, alpha)
D2 = displace(N, -alpha)
# create superposition of coherent states + random ket
psi = (D1 + D2) * basis(N,0) + 0.5 * rand_ket(N)
psi = psi.unit() # normalize
# calculate Wigner function
xvec = linspace(-5, 5, 500)
yvec = xvec
W = wigner(psi, xvec, yvec)

## new function calls ##
# store Wigner function to file
file_data_store("wigner.dat", W, numtype='real')
# load input data from file
input_data = file_data_read('wigner.dat')

# plot using mayavi
from mayavi.mlab import *
X,Y = meshgrid(xvec, yvec)
surf(xvec, yvec, input_data, warp_scale='auto')
view(0, 45)
show()
\end{verbatim}
\end{footnotesize}

\section*{Acknowledgements}
JRJ and PDN were supported by Japanese Society for the Promotion of Science (JSPS) Foreign Postdoctoral Fellowship No.~P11501 and P11202, respectively. PDN also acknowledges support from Kakenhi grant number 2301202 and Korea University.  FN acknowledges partial support from the ARO, National Science Foundation (NSF) grant No. 0726909, Grant-in-Aid for Scientific Research (S), MEXT Kakenhi on Quantum Cybernetics, and the JSPS-FIRST program.

\appendix
\section{New functions in QuTiP 2.1}\label{sec:list}
\onecolumn
\begin{table}[ht]
\begin{footnotesize}
\begin{center}
\begin{tabular}{lll}
\\ \hline\hline
\\
\textit{\small{Quantum Object Methods}} &
\\
\texttt{conj} & Conjugate of quantum object.
\\
\texttt{eigenenergies} & Calculates the eigenvalues (eigenenergies if the operator is a Hamiltonian) for a quantum operator.
\\
\texttt{groundstate} & Returns the eigenvalue and eigenstate corresponding to the ground state of the quantum operator.
\\
\texttt{matrix\_element} & Gives the matrix element $Q_{nm}=\langle \psi_{n}\left|\hat{Q}\right|\psi_{m}\rangle$ for the given operator and states $\psi_{n}$ and $\psi_{m}$.
\\
\texttt{tidyup} & Removes the small elements from a quantum object.
\\
\texttt{trans} & Transpose of a quantum object.
\\
\\
\vspace{0.5mm}
\textit{\small{Bloch-Redfield Functions}} &
\\
\texttt{bloch\_redfield\_solve} & Evolve the ODEs defined by the Bloch-Redfield tensor.
\\
\texttt{bloch\_redfield\_tensor} & Bloch-Redfield tensor for a set of system-bath operators and corresponding spectral functions.
\\
\\
\vspace{0.5mm}
\textit{\small{Floquet Functions}} &
\\
\texttt{floquet\_modes} &  Calculate the initial Floquet modes given a periodic time-dependent Hamiltonian.
\\
\texttt{floquet\_modes\_t} &  Calculate the Floquet modes at a time $t$.
\\
\texttt{floquet\_modes\_table} &  Calculate a table of Floquet modes for an interval of times.
\\
\texttt{floquet\_modes\_t\_lookup} &  Look up the Floquet modes at an arbitrary time $t$ given a pre-computed Floquet-mode table.
\\
\texttt{floquet\_states} &  Calculate the initial Floquet states given a set of Floquet modes.
\\
\texttt{floquet\_states\_t} &  Calculate the Floquet states for an arbitrary time $t$ given a set of Floquet modes.
\\
\texttt{floquet\_state\_decomposition} &  Decompose an arbitrary state vector in the basis of the given Floquet modes.
\\
\texttt{floquet\_wavefunction} &  Calculate the initial wavefunction given a Floquet-state decomposition and Floquet modes.
\\
\texttt{floquet\_wavefunction\_t} &  Calculate the wavefunction for an arbitrary time $t$ given a Floquet-state decomposition and modes.
\\
\\
\vspace{0.5mm}
\textit{\small{Evolution Solvers}} &
\\
\texttt{brmesolve} & Bloch-Redfield master equation solver.
\\
\texttt{fmmesolve} & Floquet-Markov master equation solver.
\\
\\
\vspace{0.5mm}
\textit{\small{Correlation Functions}} &
\\
\texttt{correlation} & Transient two-time correlation function.
\\
\texttt{correlation\_ss} & Steady-state two-time correlation function.
\\
\\
\vspace{0.5mm}
\textit{\small{Entropy/Entanglement Functions}} &
\\
\texttt{concurrence} & Calculate the concurrence entanglement measure for a two-qubit state.
\\
\texttt{entropy\_conditional} & The conditional entropy $S(A|B)=S(A,B)-S(B)$ of a selected density matrix component.
\\
\texttt{entropy\_mutual} & Mutual information $S(A:B)$ between selection components of a system density matrix.
\\
\\
\vspace{0.5mm}
\textit{\small{Quantum Process Tomography}} &
\\
\texttt{qpt} & Quantum process tomography $\chi$-matrix for a given (possibly non-unitary) transformation matrix U.
\\
\texttt{qpt\_plot} & Visualize the quantum process tomography $\chi$-matrix. Plot the real and imaginary parts separately.
\\
\texttt{qpt\_plot\_combined} & $\chi$-matrix plot with height and color corresponding to the absolute value and phase, respectively.
\\
\\
\vspace{0.5mm}
\textit{\small{Random State/Operator Generation}} &
\\
\texttt{rand\_dm} & Random $N\times N$ density matrix.
\\
\texttt{rand\_herm} & Random $N\times N$ Hermitian operator.
\\
\texttt{rand\_ket} & Random $N\times 1$ state (ket) vector.
\\
\texttt{rand\_unitary} & Random $N\times N$ Unitary operator.
\\
\\
\vspace{0.5mm}
\textit{\small{Gates}} &
\\
\texttt{iswap} & Quantum object representing the  $i$SWAP gate.
\\
\texttt{sqrtiswap} & Quantum object representing the square root $i$SWAP gate.
\\
\\
\vspace{0.5mm}
\textit{\small{Utility Functions}} &
\\
\texttt{file\_data\_read} & Retrieves an array of data from the requested file
\\
\texttt{file\_data\_store} & Stores a matrix of data to a file to be read by an external program.
\\
\texttt{qload} & Loads quantum object or array data contained from given file.
\\
\texttt{qsave} & Saves any quantum object or array data to the specified file.
\\
\texttt{rhs\_generate} & Pre-compiles the Cython code for time-dependent \texttt{mesolve} problems run inside a \texttt{parfor} loop.
\\
\texttt{rhs\_clear} & Clears string-format time-dependent Hamiltonian data.
\\
\hline\hline

\end{tabular}
\end{center}
\caption{List of new user-accessible functions available in QuTiP 2.1.  Additional information about each function may be obtained by calling \texttt{help(function\_name)} from the Python command line, or at the QuTiP website \cite{qutip}.}
\label{tbl:list}
\end{footnotesize}
\end{table}
\twocolumn

\bibliographystyle{model1a-num-names}
\bibliography{text}

\end{document}